\documentclass[11pt]{revtex4}

\usepackage{graphicx}
\usepackage{dcolumn}
\usepackage{bm}
\usepackage{epsfig}

\begin{document}%

\title{Modeling Crowd Turbulence by Many-Particle Simulations}
\author{Wenjian Yu and Anders Johansson}
\address{Institute for Transport \& Economics, Dresden University of Technology,
Andreas-Schubert Str. 23, 01062 Dresden, Germany
\\
Department of Humanities and Social Sciences, ETH Zurich, UNO D 11
Universit\"{a}tstrasse 41, 8092 Zurich, Switzerland }

\begin{abstract}
A recent study [D. Helbing, A. Johansson and H. Z. Al-Abideen, {\it
Phys. Rev. E} 75, 046109 (2007)] has revealed a ``turbulent" state
of pedestrian flows, which is characterized by sudden displacements
and causes the falling and trampling of people. However, turbulent
crowd motion is not reproduced well by current many-particle models
due to their insufficient representation of the local interactions
in areas of extreme densities. In this contribution, we extend the
repulsive force term of the social force model to reproduce crowd
turbulence. We perform numerical simulations of pedestrians moving
through a bottleneck area with this new model. The transitions from
laminar to stop-and-go and turbulent flows are observed. The
empirical features characterizing crowd turbulence, such as the
structure function and the probability density function of velocity
increments are reproduced well, i.e. they are well compatible with
an analysis of video data during the annual Muslim pilgrimage.
\\
\\
\end{abstract}
\maketitle

\section{Introduction}
Pedestrian dynamics\cite{crowdturbulence, PREsfm, stillPhD} has been
described by physicists through various macroscopic and microscopic
models. Macroscopic models, predominantly fluid-dynamic models
\cite{fluidRev,fluidTRB}, have the advantage of describing the
large-scale dynamics of crowds, especially depicting intermittent
flows and stop-and-go flows. These features are basically understood
as the effect of a hybrid continuity equation \cite{PRLbottleneck}
with two regimes: forward pedestrian motion and backward gap
propagation. Details of pedestrian interactions are neglected in
these models. In contrast, microscopic models can be used to
describe the details of pedestrian behavior. Previously proposed
pedestrian models include many-particle force
models\cite{PREcfm,panic}, CA models \cite{CABurs,CAMura} and
others, e.g. multi-agent approaches \cite{AgentM}, which have
received an increasing attention among physicists in the past. In
recent years, the interest has turned to empirical or experimental
studies \cite{Exp1,Exp2,Exp3,Exp4,Exp5,Exp6,transci} of pedestrian
flows based on video analysis
\cite{crowdturbulence,VideoAnalysis,Exp5,Exp2,KerridgeEmpirical}.
This has contributed to the calibration of current models
\cite{calibration,hoogendoornPED} and the discovery of new phenomena
such as crowd turbulence \cite{crowdturbulence}, which can help to
understand many crowd disasters.

Turbulent motion of pedestrians occurs, when the crowd is extremely
compressed, and people attempt to gain space by pushing others,
which leads to irregular displacements, or even the falling of
people. If the fallen pedestrians do not manage to stand up quickly
enough, they will become obstacles and cause others to fall as well.
Such dynamics can eventually spread over a large area and result in
a crowd disaster.

However, crowd turbulence is not well reproduced and understood by
pedestrian models yet, which challenges current many-particle
models. Their shortcoming is due to the underestimation of the local
interactions triggered by high densities. In the following sections,
we will extend the repulsive force of the social force model
\cite{PREsfm,panic,transci}, which has successfully depicted many
observed self-organized phenomena, such as lane formation in counter
flows and oscillatory flows at bottlenecks \cite{sfmBook}.

\section{Model of Crowd Turbulence} \label{sec_model}
Previous empirical studies\cite{crowdturbulence,Fruin} have
revealed that people are involuntarily moved when they are densely
packed, and as a consequence, the interactions increases in
areas of extreme densities, which leads to an instability of
pedestrian flows. When the average density is increasing, sudden
transitions from laminar to stop-and-go and turbulent flows are
observed. Moreover, the average flow does not reach zero.
We will now show how the turbulent flows can be modeled, by
a small extension of the social force model\cite{PREsfm,panic,transci}.
What we do is to add an extra term
to the repulsive force, and show how this will give qualitatatively different
dynamics, leading to turbulent flows.

The social force model assumes that a pedestrian $i$ tries to move
in a desired direction $\vec{e}_i$ with desired speed $v_{i}^{0}$,
and adapt the actual velocity $\vec{v}_i$ to the desired velocity
$v_i^0 \vec{e}_i$ within the relaxation time $\tau$. The velocity
$\vec{v}_i = d \vec{r}_i/dt$, i.e. the temporal change of the
position $\vec{r}_i$ is also affected by repulsive forces.

The social force model is given by
\begin{eqnarray}
m_{i}\frac{d\vec{v_{i}}(t)}{dt} = \vec{f_{i}}(t),
\end{eqnarray}
where $ \vec{f}_i(t)$ is the acceleration force of pedestrian $i$
\begin{eqnarray}
 \vec{f}_i(t) = m_{i}\frac{1}{\tau}(v_{i}^{0}\vec{e}_i -
 \vec{v}_i) + \sum_{j(\neq{i})}\vec{f}_{ij}(t).
\end{eqnarray}

The term $\vec{f}_{ij}(t)$ denotes the repulsive force, which
represents both the attempt of pedestrian $i$ to keep a certain
safety distance to other pedestrians $j$ and the desire to gain more
space in very crowded situations.

Instead of introducing an additional force term, one may reflect the
desire to gain more space under crowded conditions, by a local
interaction range in the repulsive pedestrian force $\vec{f}_{ij}$,
which is proposed as follows,

\begin{eqnarray}
\vec{ f_{ij}
}=F\Theta(\varphi_{ij})\exp[-d_{ij}/D_{0}+(D_{1}/d_{ij})^k]\vec{e}_{ij},
\end{eqnarray}

where $F$ is the maximum repulsive force ( assuming there is no
overlapping/compression ); $d_{ij}$ is the distance between center
of masses of pedestrians; $k$, $D_{0}$, and $D_{1}$ are constants;
$\vec{e}_{ij}$ is the normalized vector pointing from pedestrian $j$
to pedestrian $i$, $\varphi_{ij}$ is the angle between
$\vec{e}_{ji}$ and the desired walking direction $\vec{e}_{i}$ of
pedestrian $i$, i.e. $\cos(\varphi_{ij}) = \vec{e}_{i}\cdot
\vec{e}_{ji}$.

In normal situations, the function $\Theta(\varphi_{ij})$
reflects the fact that pedestrians react much
stronger to what happens in front of them, and it has been suggested
\cite{sfmBook} to have the form,
\begin{eqnarray}
\Theta(\varphi) = \Big(\lambda + ( 1 -
\lambda)\frac{1+\cos(\varphi)}{2}\Big).
\end{eqnarray}

Also note that when $d_{ij}$ is very small, i.e., people are squeezed,
the repulsive force will increase greatly, which reflects
the strong reactions of those located in extremely dense areas.

In the original social force model, the second repulsion term
$kg(r_{ij}-d_{ij})$ \cite{panic} reflects the physical contacts of
pedestrians, which will separate pedestrians, when collisions occur.
Here k is a constant, and function g(x) is zero, if pedestrians do
not touch each other, otherwise it is equal to the argument $x$.
In highly dense areas, the speeds of pedestrians are very low, so the small
fluctuations of $d_{ij}$ may not lead to sufficient forces for the
occurrence of turbulence, since this repulsive force is increased
linearly.
Suppose that pedestrian $i$ is located in an extremely dense area.
For the extended model, the small fluctuations of $d_{ij}$ will change the repulsive forces
greatly, and lead to sudden involuntary displacements. With the
influence of such strong reactions, the motion of pedestrians near
$i$ will be affected, and will further spread the irregular
displacements to a larger area. Thus the turbulent motion of
pedestrians will be triggered.

\section{Results and Discussion}
Numerical simulations (supplementary videos \cite{supplement}) will
now be carried out, of a crowd going through a bottleneck. The
simulations will be performed with the extended-social-force model
described in Sec.~\ref{sec_model}.

In our simulations, the desired speeds $v^{0}$ are assumed to be
Gaussian distributed with the mean value $v^{0}=1.34m/s$ and the
standard deviation $0.26m/s$ \cite{Param1,Param2,Param3}. The
relaxation time, $\tau$ is set to $0.5s$. The average mass of a
pedestrian is set to $60kg$ with a standard deviation of $10kg$.
Assuming that pedestrians have a strong
desire to gain more space in dense areas, the maximum repulsive
force $F$ is set to $160N$. Note that, when pedestrians are
compressed, the maximum force can be significantly larger than F.
The other parameters are set to, $\lambda=0.25$, $k=2$, $D_{0} =
0.31$ and $D_{1} = 0.45$.

\begin{figure}[!htbp]
\includegraphics[width=8cm, angle = 270]{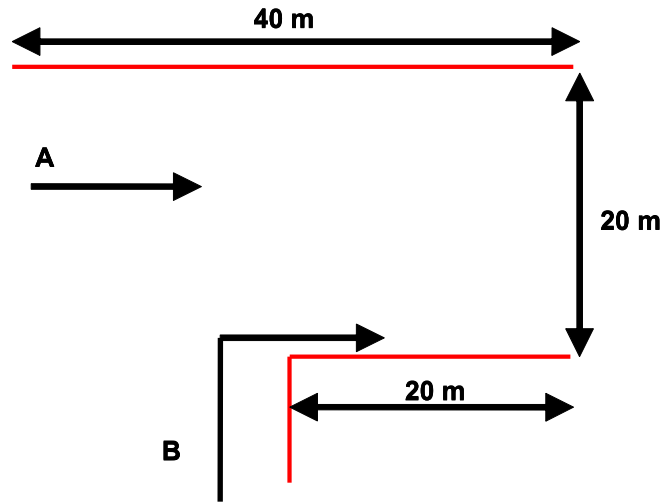}\,
\caption[]{(Color online) The bottleneck area of the simulation. Two
groups of people set out respectively from A and B, and intersect,
causing pedestrians to be highly compressed, which eventually
results in turbulent flows.} \label{fig_one}
\end{figure}

The bottleneck area (see Fig.~\ref{fig_one}) contains two pedestrian
sources, $A$ and $B$, where $A$ denotes those walking from left to
right, while $B$ represents those walking from the bottom, and then
turn right to join those coming from left. These two sources give an
increasing number of pedestrians in time, since we use both
pedestrian sorces as well as periodic boundary conditions. With this scheme we can
see how the transition from laminar to turbulent flow occur when the
density is growing. The whole area is $40m \times 20m$. The reason
for this setup, is to get a situation that is comparable to the
one in the empirical study\cite{crowdturbulence}.

The preferred direction is defined as follows{\b :}
If the vertical position of a pedestrian is above the corner, he/she
will walk in the right direction. Otherwise, he/she will first walk
upwards until he/she is above the corner, and then turn to the right
and keep walking straight ahead.

Note that interactions increase greatly when those two crowds
intersect, especially in high density situations.

\par

The ``Crowd pressure"\cite{crowdturbulence}, reflecting the
irregular/chaotic motion of people, is given by
\begin{eqnarray}
p = \rho_i \mbox{Var}(\vec{v}_i),
\end{eqnarray}
where  $\mbox{Var}(\vec{v}_i)$ is the local velocity variance. The
local density $\rho_{i}$,
\begin{eqnarray}
\rho_{i}=\sum_{j}\frac{1}{\pi
R^{2}}exp(-\|\vec{r_{j}}(t)-\vec{r_{i}}(t)\|^{2}/R^{2}).
\end{eqnarray}
$R$ is a parameter reflecting the range of smoothing. For further
aspects regarding the definition of the local density and pressure,
see Ref.\cite{crowdturbulence}.

\begin{figure}[!htbp]
\includegraphics[width=10cm]{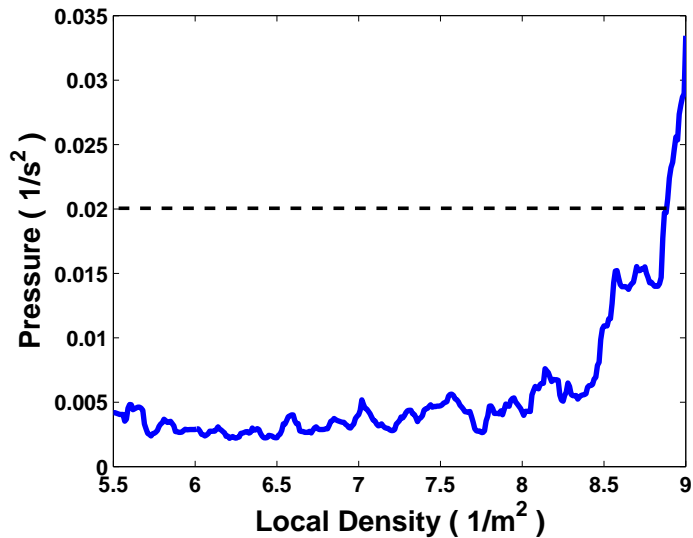}\,
\caption[]{(Color online) The crowd pressure as a function of local
density. The turbulence starts when the pressure exceeds a value of
$0.02/s^{2}$\cite{crowdturbulence}.} \label{fig_two}
\end{figure}

The pressure stays low during an increasing density, until a point
where the pressure suddenly peaks, which leads to the turbulent
crowd motion (see Fig.~\ref{fig_two}).

\par

\begin{figure}[!htbp]
\includegraphics[width=10cm]{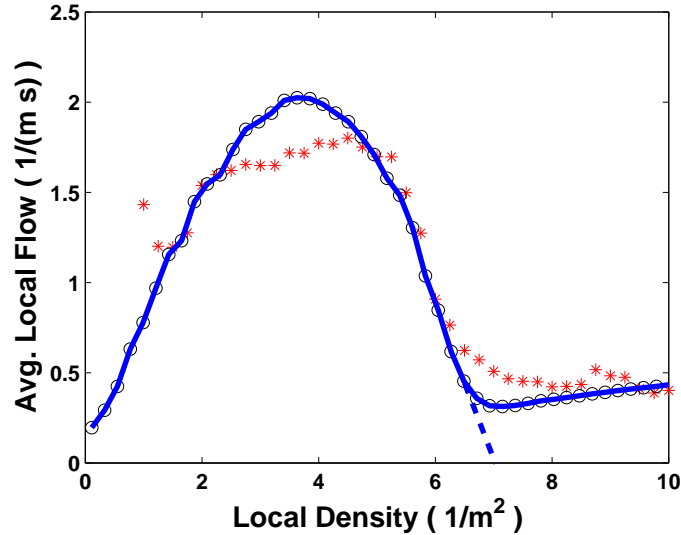}\,
\caption[]{(Color online) The average of the local flow
\cite{crowdturbulence} as a function of the local density. The
empirical data is represented by stars and the simulation results
are represented by circles connected by a solid line. The dashed
line shows the fundamental diagram, from the original social force
model.} \label{fig_three}
\end{figure}

The fundamental diagram (see Fig.~\ref{fig_three}) demonstrates the
effect of the extended repulsive force term. One can clearly see
 the difference at the tail of the curve, where the flow remains finite with the increase of local
density. Note that the flow here is not reduced to zero, even if the
density is very high due to the strong interactions within the
crowd, which will prevent people from stopping. If the flow is
reduced to zero, which means all the people stop moving, then there
is no turbulence. Therefore it is essential to have nonvanishing
flow for high densities. This is compatible with the empirical study
\cite{crowdturbulence}. Note that, at this point, the flow is no
longer laminar. Therefore, the strong interactions between
pedestrians are potentially more dangerous for the crowd. The motion
of pedestrians become turbulent, and people are pushed into all
possible directions. As people are pushed by those behind, the
fallen people will be trampled, if they do not get back on their
feet quickly enough. However, in our simulations, we assume that
pedestrians will never fall, since we are focussing on the dynamics
of the crowd during a high level of crowdedness. These conditions
can potentially lead to an accident, but we do not focus on the
dynamics of the accident itself.

\par

\begin{figure}[!htbp]
\includegraphics[width=10cm]{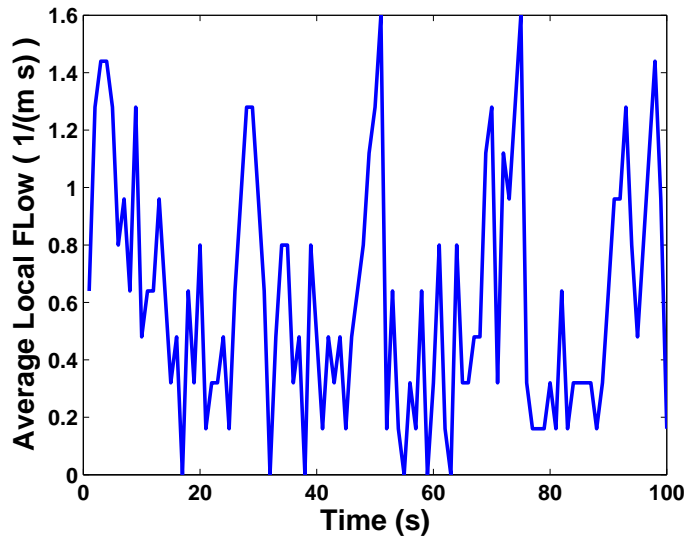}\,
\caption[]{ (Color online) Simulation results. During high crowd
densities, the smooth, laminar flow will turn into stop-and-go flow.
Here the average density is $4.3/m^2$.} \label{fig_stopandgo}
\end{figure}

With the increment of density, the pedestrian flow suddenly turns
into stop-and-go flow (see Fig.~\ref{fig_stopandgo}), which is
characterized by temporarily interrupted and longitudinally
unstable flow. This phenomenon is also predicted by a recent theoretical
approach \cite{PRLbottleneck}, which suggests that intermittent
flows at bottlenecks can be triggered when the inflow exceeds the
outflow.

\par

\begin{figure}[!htbp]
\includegraphics[width=10cm]{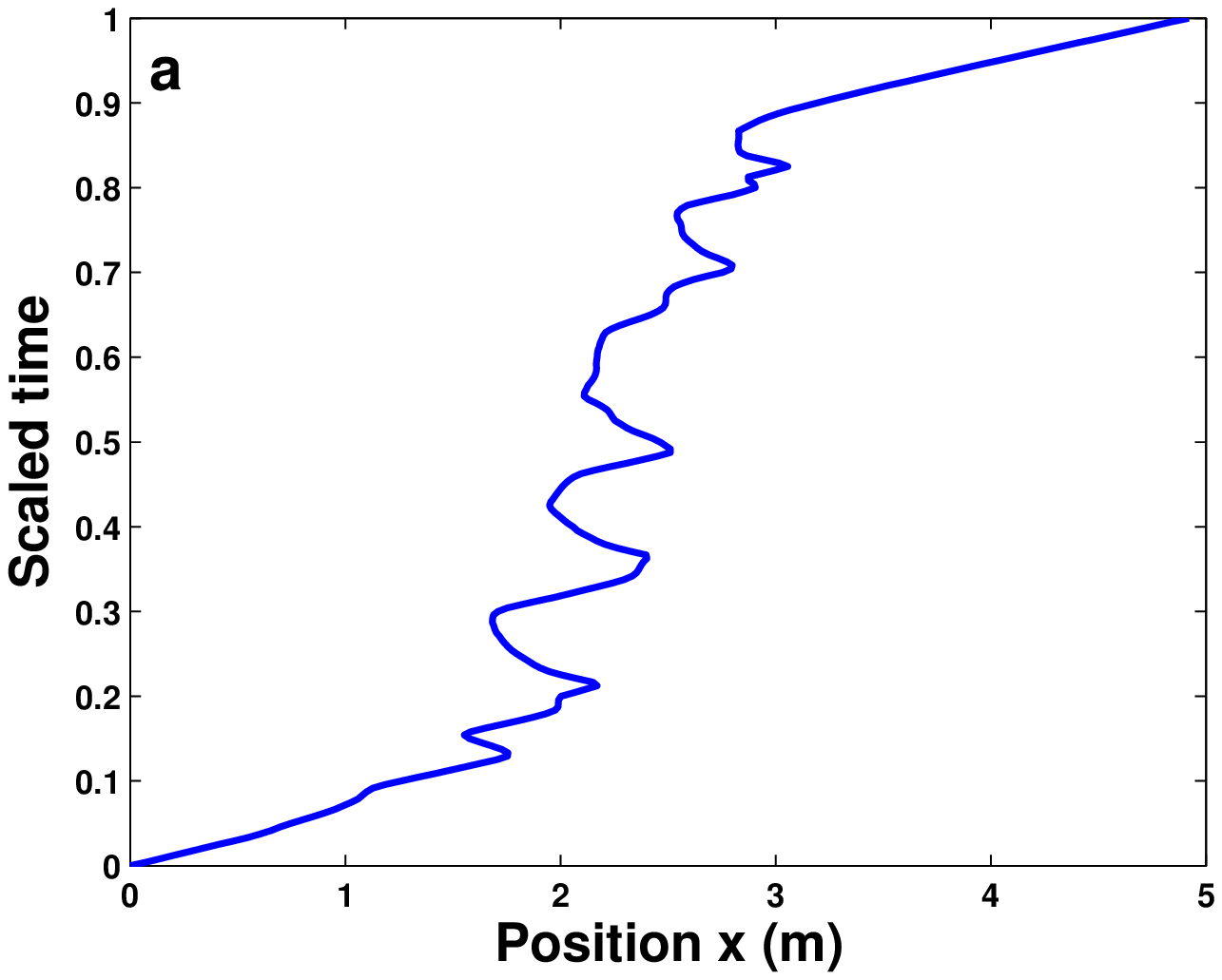}\,
\includegraphics[width=10cm]{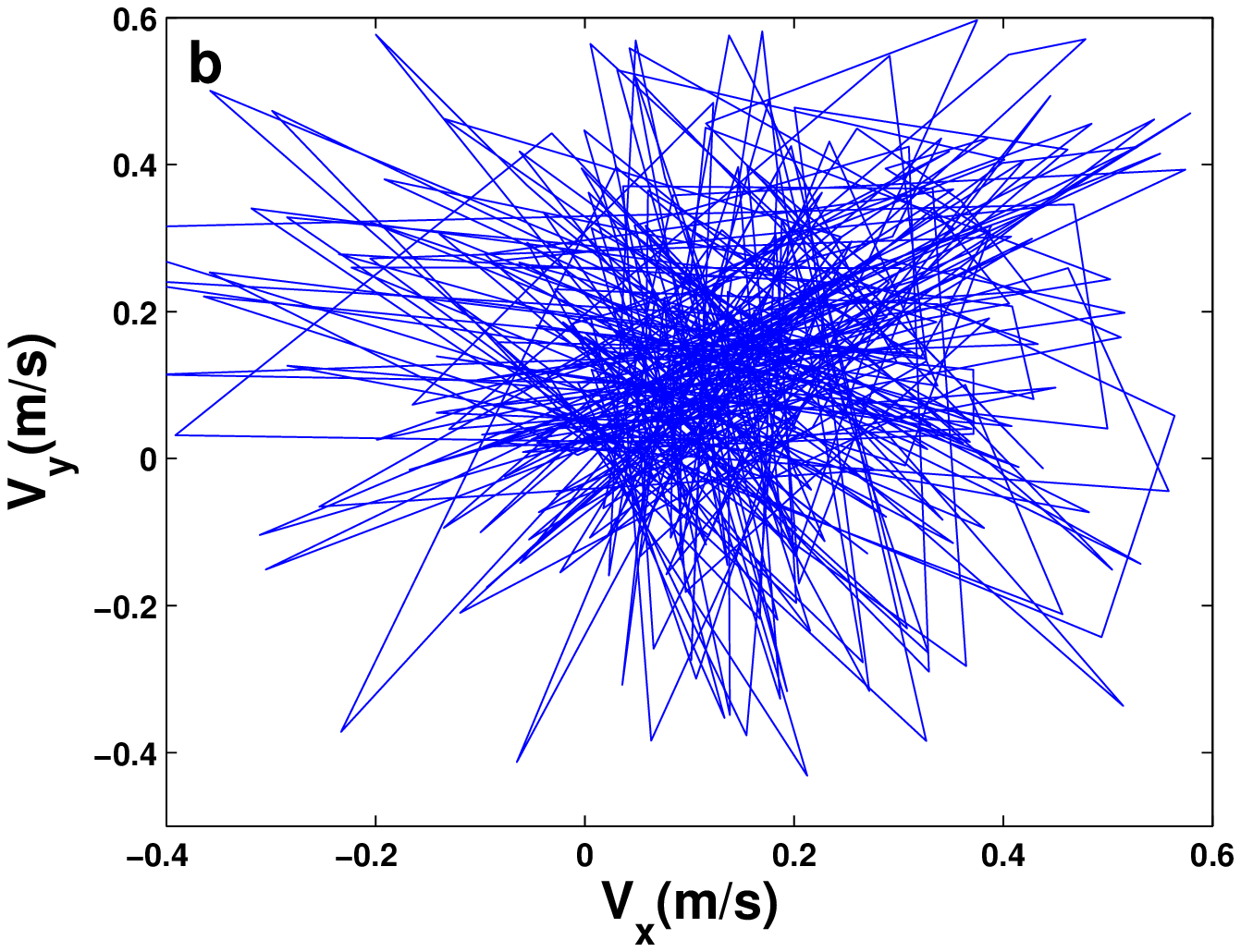}\,
\caption[]{(Color online) (a) A representative pedestrian trajectory
in laminar and turbulent flow. (b) An example of velocity components
in the turbulent motion. Here the average density is $9.0/m^2$.}
\label{fig_four}
\end{figure}

Further increment of the density will lead to turbulent flows.
Figure~\ref{fig_four}(a) shows a typical trajectory of the turbulent
motion from our simulations. We can see that, first, the curve is
smooth, which represents a laminar flow, then suddenly, vibrations
occur due to the turbulence. Also, note that the pedestrian is
sometimes even pushed backwards. The turbulent motion does not vanish
until an individual walks out of the extremely dense central area,
where the two streams intersect. Figure \ref{fig_four}(b) is an
example of the temporal evolution of an individual's velocity
components $v_{y}$ and $v_{x}$. One can clearly see the irregular
motion into all directions. Although no large eddies are observed,
as in turbulent fluids, there is still an analogy to the turbulence
of the currency exchange market \cite{exchangemarket}. This can be
characterized by the probability density function of velocity
increments and the so-called structure function,
Eq.~\ref{eq_structure}.

\par

The shape of the probability density function of velocity increment is a typical
indicator for turbulence, if the time shift $\tau'$ is small enough, and is
given by
\begin{eqnarray}
v_{x}^{\tau'} = v_{x}( \vec{r}, t+\tau' ) - v_{x}(\vec{r},t).
\end{eqnarray}

The structure function,
\begin{eqnarray}
S(\bigtriangleup \vec{r}) = \langle \|
\vec{v}(\vec{r}+\bigtriangleup\vec{r},t) - \vec{v}(\vec{r}) \|^{2}
\rangle_{\vec{r},t},
\label{eq_structure}
\end{eqnarray}
reflects the dependence of the relative speed on the distance.

\begin{figure}[!htbp]
\includegraphics[width=10cm]{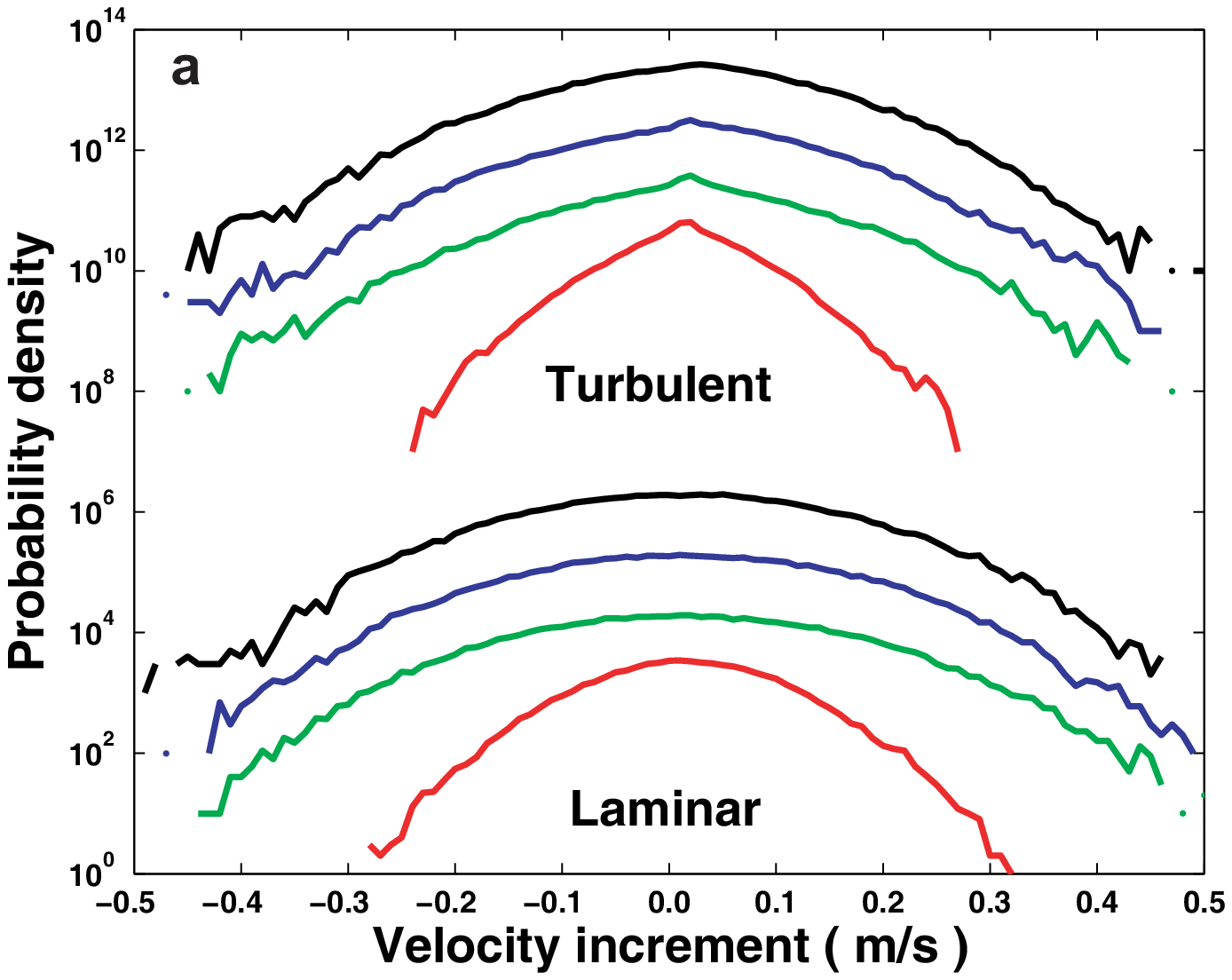}\,
\includegraphics[width=10cm]{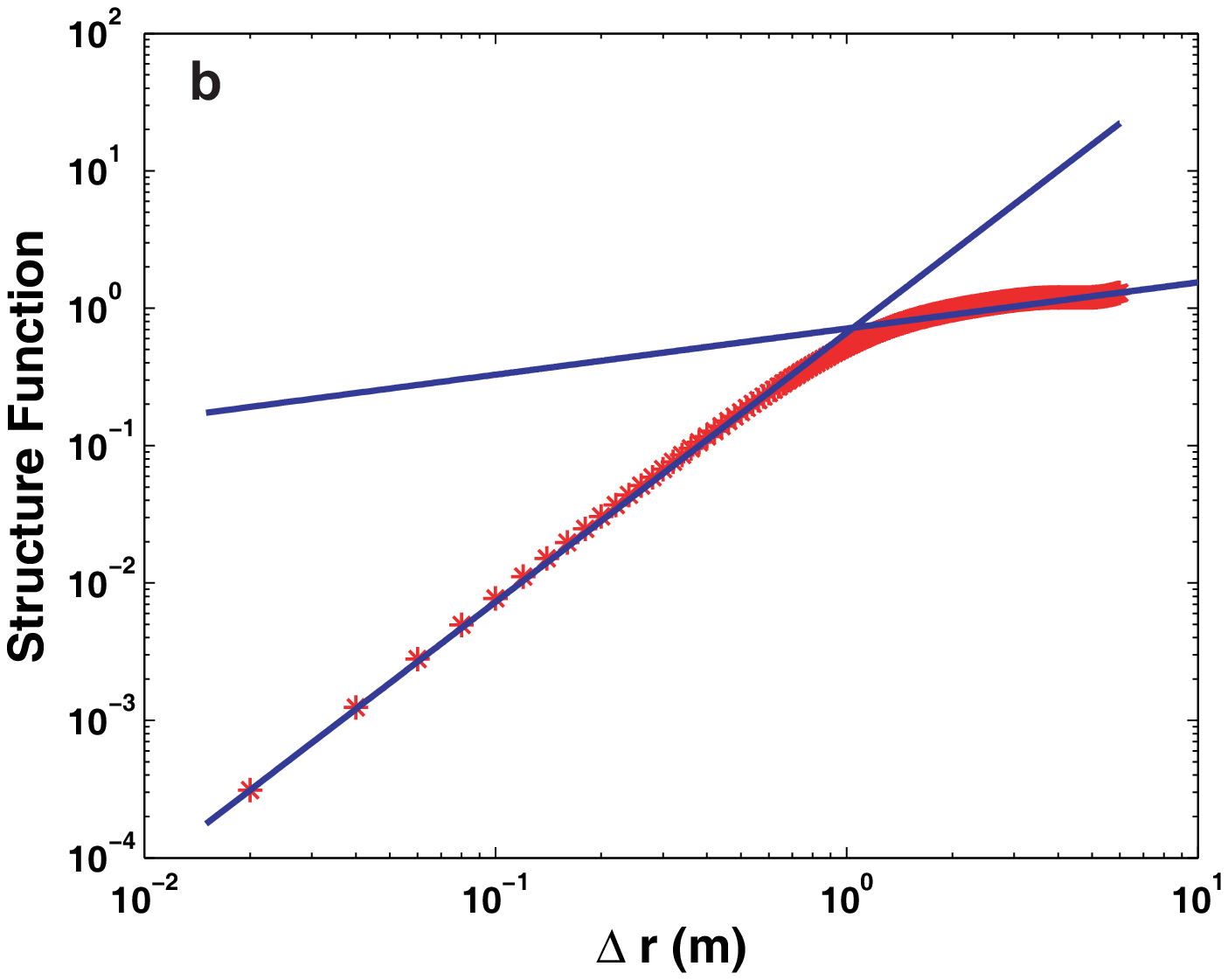}\,
\caption[]{(Color online) (a) the probability density function of
the velocity increment in the laminar and the turbulent region,
determined over many locations $\vec{r}$ for $R=\sqrt{10/\rho}$ and
$\tau' = 2s$( black curves, top ), $\tau' = 1s$( blue curves, second
), $\tau' = 0.5s$( green curves, third ), and $\tau' = 0.05s$( red
curves, bottom ). For the clarity of the presentation, the curves
are shifted in the vertical direction. The non-parabolic, peaked
curve for small values of $\tau'$ distinguishes turbulence from
laminar flows. (b) The log-log plot of the structure function for
crowd turbulence. The average density here is $9.0/m^2.$}
\label{fig_five}
\end{figure}

We find that the probability functions of velocity increment is
sharply peaked for turbulence, while it is parabolic-like for
laminar flow (see Fig.~\ref{fig_five}), when the time shift $\tau'$
is small enough. The structure function has a slope of $2.0$, when the
distance is small, while at large steps, the slope turns to $0.18$
due to the increased interactions in crowded regions. Both of
the functions are compatible to the analysis of the video recordings
of the Jamarat Bridge\cite{crowdturbulence}.

\par

\section{Summary and Outlook}
In summary, an added social force component, reflecting the strong
interactions in the extremely crowded areas is proposed for the
simulation of crowd turbulence, which questions current
many-particle models. The transition from laminar to stop-and-go and
turbulent flows is observed in the simulations. The fundamental
diagram is reproduced and demonstrates the effects of the extended
repulsive forces at highly dense situations, i.e. the average local
flow is not reduced to zero. A typical turbulent trajectory and
velocity components are presented. Functions like the probability
density function of velocity increment and the structure function,
characterizing the features of turbulence are simulated, and the
results are compatible to the empirical studies.

\begin{acknowledgments}
The authors are grateful to the German Research Foundation for
funding ( DFG project He 2789/7-1 ).
\end{acknowledgments}

\end{document}